*Proceeding Paper*

# Conformal Symmetries of the Strumia–Tetradis' Metric †

**Pantelis S. Apostolopoulos** [1,*] **and Christos Tsipogiannis**

Department of Environment, Mathematical Physics and Computational Statistics Research Laboratory, Ionian University, 29100 Zakynthos, Greece
* Correspondence: papostol@ionio.gr
† Presented at the 2nd Electronic Conference on Universe, 16 February–2 March 2023; Available online: https://ecu2023.sciforum.net/.

**Abstract:** In a recent paper, a new conformally flat metric was introduced, describing an expanding scalar field in a spherically symmetric geometry. The spacetime can be interpreted as a Schwarzschild-like model with an apparent horizon surrounding the curvature singularity. For the above metric, we present the complete conformal Lie algebra consisting of a six-dimensional subalgebra of isometries (Killing Vector Fields or KVFs) and nine proper conformal vector fields (CVFs). An interesting aspect of our findings is that there exists a gradient (proper) conformal symmetry (i.e., its bivector $F_{ab}$ vanishes) which verifies the importance of gradient symmetries in constructing viable cosmological models. In addition, the 9-dimensional conformal algebra implies the existence of constants of motion along null geodesics that allow us to determine the complete solution of null geodesic equations.

## 1. Introduction

From the era of the development of differential geometry as an inherent constituent of the spacetime continuous, it became clear that certain important classes of spaces have remarkable properties. Based on these properties, these spaces can be classified independently of the coordinate system used, or the physical system that they describe. These properties are usually called symmetries and can be divided into three basic categories depending on the nature of the object they act upon: geometric symmetries (also known as collineations) refer to the set of spacetime symmetries that are related to the geometric structure of the model; kinematic symmetries refer to the set of spacetime symmetries that interact with the kinematic quantities; and matter symmetries that act on the dynamic quantities of the model. Furthermore, it was confirmed in later studies that, among other things, these properties have a direct physical interpretation. A characteristic example is the topology of the $n$-dimensional sphere, which has constant and positive curvature and is characterized by three principal directions that also constitute its symmetry axes. Geometrically, this means that the inner products remain constant, so the metric of the sphere remains unchanged along the integral curves of the principal directions that determine its symmetry axes. In addition, like any space of maximum symmetry, it admits a Lie algebra of conformal symmetries, which, in combination with the algebra of isometries, is used for the complete description of the sphere in terms of collineations. The concept of conformal symmetries is an important tool for understanding the geometric structure of non-linear spaces, such as manifolds, and their applications in physics. In fact, sufficient geometries that are widely used in various physical theories are intrinsically characterized by the existence of conformal (proper or not) symmetries (e.g., [1–8]).

In the present article, the spacetime signature is assumed to be $(-,+,+,+)$, with lowercase Latin letters denoting spacetime indices $a,b,\ldots = 0,1,2,3$ and we use geometrized units such that $8\pi G = c = 1$.

## 2. Methods

In a recent paper [9], a new conformally flat metric was found, describing an expanding scalar bubble within a spherically symmetric geometry having a Schwarzschild-like behaviour. The solution (up to some integration constants) is:

$$\mathsf{c}(t,r) = \left[1 - \frac{c_1^2}{\left(1 + \frac{r^2 - t^2}{c_2^2}\right)^2}\right]^{1/2} \tag{1}$$

where the metric of the spacetime is:

$$ds^2 = \mathsf{c}(t,r)^2 \left[-dt^2 + dr^2 + r^2\left(d\vartheta^2 + \sin^2\vartheta d\phi^2\right)\right]. \tag{2}$$

It is observed that the flat part of the metric (2):

$$ds^2_{\text{FLAT}} = -d\tau^2 + dx^2 + dy^2 + dz^2 \tag{3}$$

has been expressed in a spherically symmetric form and therefore, it will be convenient to locate the transformation between Cartesian and spherical coordinates. In addition, the full group of conformal symmetries of (4) is represented with the vectors ($\mathbf{X}_1 - \mathbf{X}_4$ are spatial and temporal translations, $\mathbf{X}_5 - \mathbf{X}_7$ are the space rotations, $\mathbf{X}_8 - \mathbf{X}_{10}$ represent spacetime rotations, $\mathbf{X}_{11}$ is the generator of self-similarity and the vectors, $\mathbf{X}_{12} - \mathbf{X}_{15}$ are the special conformal Killing vectors) [10]:

$$\begin{aligned}
&\mathbf{X}_1 = \partial_\tau, \quad \mathbf{X}_2 = \partial_x, \quad \mathbf{X}_3 = \partial_y, \quad \mathbf{X}_4 = \partial_z \\
&\mathbf{X}_5 = -y\partial_x + x\partial_y, \quad \mathbf{X}_6 = z\partial_x - x\partial_z, \quad \mathbf{X}_7 = -z\partial_y + y\partial_z \\
&\mathbf{X}_8 = x\partial_\tau + \tau\partial_x, \quad \mathbf{X}_9 = y\partial_\tau + \tau\partial_y, \quad \mathbf{X}_{10} = z\partial_\tau + \tau\partial_z \\
&\mathbf{X}_{11} = \tau\partial_\tau + x\partial_x + y\partial_y + z\partial_z \\
&\mathbf{X}_{12} = \left(\tau^2 + x^2 + y^2 + z^2\right)\partial_\tau + 2\tau x\partial_x + 2\tau y\partial_y + 2\tau z\partial_z \\
&\mathbf{X}_{13} = 2\tau x\partial_\tau + \left(\tau^2 + x^2 - y^2 - z^2\right)\partial_x + 2xy\partial_y + 2xz\partial_z \\
&\mathbf{X}_{14} = 2\tau y\partial_\tau + 2xy\partial_x + \left(\tau^2 + y^2 - x^2 - z^2\right)\partial_y + 2yz\partial_z \\
&\mathbf{X}_{15} = 2\tau z\partial_\tau + 2xz\partial_x + 2yz\partial_y + \left(\tau^2 + z^2 - x^2 - y^2\right)\partial_z.
\end{aligned} \tag{4}$$

Using the coordinate transformation $(\tau, x, y, z) \to (t, r, \phi, \vartheta)$:

$$\begin{aligned}
&\tau(t,r,\phi,\vartheta) = t, \quad x(t,r,\phi,\vartheta) = r\sin\phi\sin\vartheta \\
&y(t,r,\phi,\vartheta) = r\cos\phi\sin\vartheta, \quad z(t,r,\phi,\vartheta) = r\cos\vartheta
\end{aligned} \tag{5}$$

we obtain:

$$ds^2_{\text{FLAT}} = -dt^2 + dr^2 + r^2\left(d\vartheta^2 + \sin^2\vartheta d\phi^2\right) \tag{6}$$

and the conformal vector fields become:

$$\mathbf{X}_1 = \partial_t, \quad \mathbf{X}_2 = \sin\vartheta\sin\phi\,\partial_r + \frac{\cos\vartheta\sin\phi}{r}\partial_\vartheta + \frac{\cos\phi}{\sin\vartheta}\partial_\phi,$$

$$\mathbf{X}_3 = \sin\vartheta\cos\phi\,\partial_r + \frac{\cos\vartheta\cos\phi}{r}\partial_\vartheta - \frac{\sin\phi}{\sin\vartheta}\partial_\phi$$

$$\mathbf{X}_4 = \cos\vartheta\,\partial_r - \frac{\sin\vartheta}{r}\partial_\vartheta$$

$$\mathbf{X}_5 = -\partial_\phi, \quad \mathbf{X}_6 = -\cos\phi\,\partial_\vartheta + \cot\vartheta\sin\phi\,\partial_\phi$$

$$\mathbf{X}_7 = \sin\phi\,\partial_\vartheta + \cot\vartheta\cos\phi\,\partial_\phi$$

$$\mathbf{X}_8 = r\sin\vartheta\sin\phi\,\partial_t + t\sin\vartheta\sin\phi\,\partial_r + \frac{t\cos\vartheta\sin\phi}{r}\partial_\vartheta + \frac{t\cos\phi}{r\sin\vartheta}\partial_\phi$$

$$\mathbf{X}_9 = r\sin\vartheta\cos\phi\,\partial_t + t\sin\vartheta\cos\phi\,\partial_r + \frac{t\cos\vartheta\cos\phi}{r}\partial_\vartheta - \frac{t\sin\phi}{r\sin\vartheta}\partial_\phi \qquad (7)$$

$$\mathbf{X}_{10} = r\cos\vartheta\,\partial_t + t\cos\vartheta\,\partial_r - \frac{t\sin\vartheta}{r}\partial_\vartheta$$

$$\mathbf{X}_{11} = t\partial_t + r\partial_r$$

$$\mathbf{X}_{12} = (r^2 + t^2)\partial_t + 2tr\,\partial_r$$

$$\mathbf{X}_{13} = 2rt\sin\vartheta\sin\phi\,\partial_t + (r^2+t^2)\sin\vartheta\sin\phi\,\partial_r + \frac{(t^2-r^2)\cos\vartheta\sin\phi}{r}\partial_\vartheta + \frac{(t^2-r^2)\cos\phi}{r\sin\vartheta}\partial_\phi$$

$$\mathbf{X}_{14} = 2rt\sin\vartheta\cos\phi\,\partial_t + (r^2+t^2)\sin\vartheta\cos\phi\,\partial_r + \frac{(t^2-r^2)\cos\vartheta\cos\phi}{r}\partial_\vartheta + \frac{(r^2-t^2)\sin\phi}{r\sin\vartheta}\partial_\phi$$

$$\mathbf{X}_{15} = 2rt\cos\vartheta\,\partial_t + (r^2+t^2)\cos\vartheta\,\partial_r + \frac{(r^2-t^2)\sin\vartheta}{r}\partial_\vartheta.$$

The conformal vector fields of the metric (2) are also given by $\mathbf{X}_1 - \mathbf{X}_{15}$ with conformal factors derived from the relation:

$$\mathbf{L}_\mathbf{X} g_{ab} = \mathbf{L}_\mathbf{X}(C^2 \eta_{ab}) = 2[\mathbf{X}(\ln C) + \Psi]g_{ab} \qquad (8)$$

where $\eta_{ab}, \Psi$ is the metric and the conformal factors of the Minkowski spacetime, respectively.

Using equation (8), we can straightforwardly determine the conformal factors of the metric (2):

$$\psi(\mathbf{X}_1) = (\ln C)_{,t} \quad \psi(\mathbf{X}_2) = \sin\vartheta\sin\phi(\ln C)_{,t} \quad \psi(\mathbf{X}_3) = \sin\vartheta\cos\phi(\ln C)_{,t}$$
$$\psi(\mathbf{X}_4) = \cos\vartheta(\ln C)_{,t} \quad \psi(\mathbf{X}_5) = \psi(\mathbf{X}_6) = \psi(\mathbf{X}_7) = 0$$
$$\psi(\mathbf{X}_8) = \sin\vartheta\sin\phi[t(\ln C)_{,r} + r(\ln C)_{,t}] \quad \psi(\mathbf{X}_9) = \sin\vartheta\cos\phi[t(\ln C)_{,r} + r(\ln C)_{,t}]$$
$$\psi(\mathbf{X}_{10}) = \cos\vartheta[t(\ln C)_{,r} + r(\ln C)_{,t}] \quad \psi(\mathbf{X}_{11}) = 1 + t(\ln C)_{,t} + r(\ln C)_{,r} \qquad (9)$$
$$\psi(\mathbf{X}_{12}) = 2t + 2tr(\ln C)_{,r} + (r^2+t^2)(\ln C)_{,t}$$
$$\psi(\mathbf{X}_{13}) = \sin\vartheta\sin\phi[2r + 2tr(\ln C)_{,t} + (r^2+t^2)(\ln C)_{,r}]$$
$$\psi(\mathbf{X}_{14}) = \sin\vartheta\cos\phi[2r + 2tr(\ln C)_{,t} + (r^2+t^2)(\ln C)_{,r}]$$
$$\psi(\mathbf{X}_{15}) = \cos\vartheta[2r + 2tr(\ln C)_{,t} + (r^2+t^2)(\ln C)_{,r}]$$

It is easily verified from the expressions (7), that the vectors $\mathbf{X}_8, \mathbf{X}_9, \mathbf{X}_{10}$ are reduced to isometries for the line element (2), with metric function $C(t,r)$ given in (1), and represent *space-time boosts* (the same is applied for the metric found in [11]). Note also that the 9-dimensional Lie Algebra of the *proper* conformal symmetries given above can be used, in principle, to determine the general solution of the null geodesic equation. In fact, the existence of a proper conformal vector field $\mathbf{X}$ implies that there is a constant of motion along null geodesics ($n^a n_a = 0$, $n_{a;b} n^b = 0$) [3]:

$$\left(X_a n^a\right)_{;b} n^b = X_{a;b} n^a n^b + X_a n^a_{;b} n^b = \psi g_{ab} n^a n^b = 0. \tag{10}$$

## 3. Results and Discussion

The spacetime (1)–(2) is a solution of the field equations with a minimally coupled scalar field. It is straightforward to see that the CVF $\mathbf{X}_{11} = t\partial_t + r\partial_r$ is a gradient (proper) conformal symmetry (i.e., its bivector $F_{ab}$ vanishes) that verifies the importance of gradient symmetries in constructing viable cosmological models. Our findings also indicate an eventually close connection between these classes of models and the existence of a gradient CVF, that has so far been underestimated.

**Funding:** This research received no external funding.

**Institutional Review Board Statement:** Not applicable.

**Informed Consent Statement:** Not applicable.

**Data Availability Statement:** Not applicable here.

**Conflicts of Interest:** : The authors declare no conflict of interest.